\documentclass[prb,nofootinbib,twocolumn,superscriptaddress]{revtex4} 

\usepackage{graphicx}
\usepackage{dcolumn}
\usepackage{bm}
\usepackage{threeparttable}
\usepackage{times}
\usepackage{mathptmx}
\usepackage{lscape}
\usepackage{natbib}
\usepackage{amsmath}
\usepackage{amssymb}
\usepackage{braket}
\usepackage{comment}
\usepackage{color}

\def\degree{\kern-.2em\r{}\kern-.3em}

\providecommand{\norm}[1]{\lVert#1\rVert}

\begin{document}


\title{  Modified Representation of Canonical Average by Special Microscopic States\\ for Classical Discrete Systems   }

\author{Koretaka Yuge}
\affiliation{
Department of Materials Science and Engineering,  Kyoto University, Sakyo, Kyoto 606-8501, Japan\\
}%

\author{Shouno Ohta}
\affiliation{
Department of Materials Science and Engineering,  Kyoto University, Sakyo, Kyoto 606-8501, Japan\\
}%

\author{Ryogo Miyake}
\affiliation{
Department of Materials Science and Engineering,  Kyoto University, Sakyo, Kyoto 606-8501, Japan\\
}%

\begin{abstract}
{
For substitutional crystalline solids typically referred to classical discrete system under constant composition, macroscopic structure in thermodynamically equilibrium state can be typically obtained through canonical average, where a set of microscopic structure dominantly contributing to the average should depend on temperature and many-body interaction through Boltzmann factor, $\exp\left(-\beta E\right)$. Despite these facts, our recent study reveals that based on configurational geometry, a few specially-selected microscopic structure (called ``projection state'': PS)  \textit{independent} of temperature and many-body interaction can reasonably characterize temperature dependence of macroscopic structure. Here we further modify representation of canonical average by using the same PSs, based on (i) transformation of multivariate 3-order moment matrix by one of the PS, and (ii) Pade approximation. We prove that the former can always results in better representation of canonical average than non-transformation one, confirmed by performing hypershere integration, while the latter approximation can provide better representaion except for e.g., inclusion of its own singular point within considered temperature, which can be known \textit{a priori}. 
  }
\end{abstract}


\maketitle

\section{Introduction}
Statistical mechanics provides that for classical discrete systems under constant composition, structure along chosen coordination $Q_{r}$  on $f$-dimensional configuration space in thermodynamically equilibrium state can be typically given by the so-called canonical average:
\begin{eqnarray}
\label{eq:can}
\Braket{Q_{r}}_{Z} \left( T \right) = Z^{-1} \sum_{d} q_{r}^{\left( d \right)} \exp\left( -\beta U^{\left( d \right)} \right),
\end{eqnarray}
where $Z$ denotes partition function, $\beta$ inverse temperature, and summation is taken over all possible microscopic states on configuration space. 
Eq.~\eqref{eq:can} clearly indicates that a set of microscopic state dominantly contributing to canonical average of the left-hand side should in principle depend on temperature and interaction, which cannot be known \textit{a priori}. 
Therefore, a variety of theoretical approaches have been amply developed to effectively sample significant microscopic states for the canonical average, including Metropolis algorism, entropic sampling and Wang-Landau method.\cite{mc1,mc2,mc3,wl} 
Despite these facts, our recent investigations find that Eq.~\eqref{eq:can} can be approximated as\cite{em1,em2} 
\begin{eqnarray}
\label{eq:emrs}
\Braket{Q_{r}}_{Z} \simeq \Braket{q_{r}} -  \sqrt{\frac{\pi}{2}}\Braket{q_{r}}_{2}\beta E_{r} + \frac{\beta^{2}}{2}\sum_{i=1}^{g} \omega_{i}E_{r_{i}}^{2},
\end{eqnarray}
where $\Braket{\quad}$ denotes taking linear average over CDOS \textit{before} applying many-body interaction to the system, $\Braket{\quad}_{2}$ denotes taking standard deviation for the CDOS,  and $\Braket{\quad|\quad}$ denotes inner product, i.e., trace over configuration space. 
$g\le f$, and $E_{r}$ and $E_{r_{2}}$ are energy of specially selected states (projections states of PS0 and PS1), whose structure also depends only on configurational geometry, and $\omega=\pm 1$ described later.
$E_{r}$ is explicitly given by\cite{em0}
\begin{eqnarray}
E_{r} = \sum_{i=1}^{g} \Braket{E|q_{i}} \Braket{q_{i}}_{r}^{\left(+\right)},
\end{eqnarray}
where $\Braket{\quad}_{r}^{\left(+\right)}$ denotes taking linear average for all possible microscopic states satisfying $q_{r} \ge \Braket{q_{r}}$.
These means that we can \textit{a priori} know structure of PSs without any thermodynamic information. 
In our previous study, we qualitatively see that for $g=1$, deviation of Eq.~\eqref{eq:emrs} from the case for $g=f$ depends on given interaction, quantitative forumation for the deviation, and corresponding modification of formulation has not been investigated. 
Since the third term of r.h.s. of Eq.~\eqref{eq:emrs} corresponds to including multivariate 3-order moments for configurational density of states (CDOS), summation up to $g=1$ corresponds to how much information of the 3-order moments can be included in a single microscopic state, $r_{1}$. 
The present study tuckle this problem, by exactly formulating the devitaion in Eq.~\eqref{eq:emrs}, and developing modified representation for $g=1$ that \textit{always} results in better approximation than Eq.~\eqref{eq:emrs} for any configurational geometry. 
We futher modify Eq.~\eqref{eq:emrs} by applying Pade approximation using the same energy information for PSs, which widely results in better representation for canonical average. 
The details are shown below.

\section{Derivation and Discussions}
\subsection{Modification with Singular Value Decomposition}
To modify Eq.~\eqref{eq:emrs} under the condition of $g=1$, we first briefly show how the third term of Eq.~\eqref{eq:emrs} is derived. 
Since we have shown that PS0 (i.e., energy of $E_{r}$) contains information about even-order multivariate moments for CDOS, the first-order correction to PS0 can be naturally the lowest odd-order moments, namely
\begin{eqnarray}
\label{eq:3mom}
\Braket{Q_{r}}_{Z}  \simeq \Braket{q_{r}} -  \sqrt{\frac{\pi}{2}}\Braket{q_{r}}_{2}\beta E_{r}+ \frac{\beta^{2}}{2}\sum_{j,k} \Braket{q_{r}q_{j}q_{k}} \Braket{E|q_{j}}\Braket{E|q_{k}}. \nonumber \\
\quad
\end{eqnarray}
To represent the third term of Eq.~\eqref{eq:3mom} by the sum of squared energy for a set of microscopic state whose structure can be known without any thermodynamic information, we previously perform singular value decomposition (SVD), leading to
\begin{eqnarray}
\label{eq:svd}
&&\sum_{j,k} \Braket{q_{r}q_{j}q_{k}} \Braket{E|q_{j}}\Braket{E|q_{k}} = {}^{t}\mathbf{X}\mathbf{A}\mathbf{X}=\sum_{i=1}^{f} \omega_{i}{}^{t}\mathbf{X}\lambda_{i} \left(\mathbf{U}_{i} \otimes \mathbf{U}_{i}\right) \mathbf{X} \nonumber \\
&&= \sum_{i=1}^{f}\omega_{i} \left\{{}^{t}\mathbf{X}\left(\lambda_{i}^{\frac{1}{2}}\mathbf{U}_{i}\right)  \right\}   \left\{ \left(\lambda_{i}^{\frac{1}{2}}\mathbf{U}_{i}^{t}\right)\mathbf{X} \right\} \nonumber \\
&&=\sum_{i=1}^{f} \omega_{i}\left\{ \sum_{m=1}^{f} \Braket{E|q_{m}} \left(\lambda_{i}^{\frac{1}{2}} U_{im} \right)  \right\}^{2} = \sum_{i=1}^{f} \omega_{i}E_{r_{i}}^{2},
\end{eqnarray}
where $X$ and $\mathbf{A}$ are $f$-dimentional vector and $f\times f$ real symmetric matrix, namely ${}^{t}X_{k}=\Braket{E|q_{k}}$ and $A_{ij}=\Braket{q_{r}q_{i}q_{j}}$, $\lambda_{i}$ and $\mathbf{U}_{i}$ are $i$-th singular value and singular vector (where $\lambda_{1}\ge \lambda_{2}\ge\cdots \ge \lambda_{f}$), $\otimes$ denotes Kronecker product, $\omega_{i}=1$ (-1) when $i$-th eigenvalue of $\mathbf{A}$ is positive (negative), and $U_{im}$ denotes $m$-th component of $i$-th singular vector. 
Since $\lambda_{i}U_{im}$ depends only on configurational geometry, and energy for given structure of $\left\{\xi_{1},\cdots,\xi_{f}\right\}$ is exactly given by
\begin{eqnarray}
E = \sum_{i=1}^{f} \Braket{E|\xi_{i}} \xi_{i},
\end{eqnarray}
$l$-th projection state $r_{l}$ have structure of $\left\{\lambda_{l}^{\frac{1}{2}}U_{l1},\cdots, \lambda_{l}^{\frac{1}{2}}U_{lf}\right\}$, which can be known \textit{a priori} without any thermodynamic information. 
Therefore, approximating the third term of Eq.~\eqref{eq:emrs} up to $g=1$ corresponds to considering the PS2 of $r_{1}$ with largest singular value, $\lambda_{1}$. 
Then, the problem to simply perform SVD to obtain the PS2 is that while SVD provides minimized Frobenius norm of
\begin{eqnarray}
\label{eq:norm}
\norm{\mathbf{A}\mp\lambda_{1}\mathbf{U}_{1}\otimes\mathbf{U}_{1}}_{\mathrm{F}} = \mathrm{min},
\end{eqnarray}
Eq.~\eqref{eq:norm}  does not generally guarantee the minimization of defferences in the third term between $g=1$ and $g=f$, namely
\begin{eqnarray}
\label{eq:F}
\left|\sum_{j,k} \Braket{q_{r}q_{j}q_{k}} \Braket{E|q_{j}}\Braket{E|q_{k}} \mp{}^{t}\mathbf{X}\lambda_{1}\left(\mathbf{U}_{1} \otimes \mathbf{U}_{1}\right)\mathbf{X}  \right|\neq\mathrm{min}. \nonumber \\
\quad
\end{eqnarray}
Here, upper-side sign (minus) takes for $\omega_{1}>0$ and lower-side (plus) for $\omega_{1}<0$ and hereinafter. 
We can clearly see that minizing the l.h.s. of Eq.~\eqref{eq:F} for any given potential energy surface: PES (i.e., any set of $\left\{\Braket{E|q_{k}}\right\}$ by fixing $\lambda_{1}$ and $\mathbf{U}_{1}$ is generally impossible. 
Therefore, we consider minimization for Eq.~\eqref{eq:F} averaged over all possible PES. 
Since Eq.~\eqref{eq:svd} corresponds to a quadratic form, it is sufficient to consider a set of PES on $S^{f-1}$ hyperspheric surface with unit radius, namely satisfying
\begin{eqnarray}
\sqrt{\sum_{i=1}^{f} \Braket{E|q_{i}}^{2} } = 1.
\end{eqnarray}
We can therefore consider root mean square (RMS) for Eq.~\eqref{eq:F} on $S^{f-1}$, given by
\begin{widetext}
\begin{eqnarray}
\label{eq:R}
R &=& \sqrt{ \int_{S^{f-1}} \left\{ \sum_{j,k} \Braket{q_{r}q_{j}q_{k}} \Braket{E|q_{j}}\Braket{E|q_{k}} \mp {}^{t}\mathbf{X}\lambda_{1}\left(\mathbf{U}_{1} \otimes \mathbf{U}_{1}\right)\mathbf{X} \right\} d\mathbf{X} \bigg/ \int_{S^{f-1}} d\mathbf{X}    } \nonumber \\
&=&  \sqrt{\frac{ \left( \sum_{i=1}^{f} v_{i}^{2} \mp A_{ii}  \right)^{2}  + 2\norm{\mathbf{A}\mp\mathbf{U}_{1}\otimes\mathbf{U}_{1}}_{\mathrm{F}}   }{f\left(f+2\right)}}   ,
\end{eqnarray}
\end{widetext}
where
\begin{eqnarray}
v_{i} = \lambda_{1}^{\frac{1}{2}} U_{1i}.
\end{eqnarray}
Equation~\eqref{eq:R} clearly supports relationships between Eq.~\eqref{eq:norm} and Eq.~\eqref{eq:F} due to the additional term of $ \left( \sum_{i=1}^{f} v_{i}^{2} \mp A_{ii}  \right)^{2}$. 
From Eq.~\eqref{eq:R}, we first examine whether or not we can obtain an optimal set of $\left\{v_{1},\cdots,v_{f}\right\}$ to minimize $R$ based on SVD of a certain matrix. 
When we define $\mathbf{V}=\mathbf{U}_{1}\otimes\mathbf{U}_{1}$, numerator of the final equation in Eq.~\eqref{eq:R} is transformed into:
\begin{widetext}
\begin{eqnarray}
\label{eq:mat}
&&\left( \sum_{i=1}^{f} v_{i}^{2} \mp A_{ii}  \right)^{2}  + 2\norm{\mathbf{A}\mp\mathbf{U}_{1}\otimes\mathbf{U}_{1}}_{\mathrm{F}} = \left\{\mathrm{Tr}\left(\mathbf{V}\mp\mathbf{A}\right)  \right\}^{2} + 2\norm{\mathbf{V} \mp \mathbf{A}}_{\mathrm{F}}^{2} = \mathrm{Tr}\left[\left(\mathbf{V}\mp\mathbf{A}\right)\otimes \left(\mathbf{V}\mp\mathbf{A}\right) \right] + 2 \mathrm{Tr} \left[ \left(\mathbf{V} \mp\mathbf{A}\right)^{2} \right] \nonumber \\
&&= \mathrm{Tr} \left[  \left(\mathbf{V}\mp\mathbf{A}\right)\otimes \left(\mathbf{V}\mp\mathbf{A}\right) + 2\mathbf{J}\otimes \left(\mathbf{V}\mp\mathbf{A}\right)^{2}   \right] = \mathrm{Tr} \left\{ \left(\mathbf{Y}\mathbf{F^{\frac{1}{2}}}\mathbf{Y}^{-1}\right)^{2} \right\} = \norm{\mathbf{F}^{\frac{1}{2}}}_{\mathrm{F}}^{2},
\end{eqnarray}
\end{widetext}
where we introduce $f\times f$ matrix of $\mathbf{J}$ given by
\begin{eqnarray}
J_{ik} &=& 1\quad\left(i=k=1\right) \nonumber \\
J_{ik} &=& 0\quad\left(\mathrm{otherwise}\right),
\end{eqnarray}
$\mathbf{Y}$ denotes $f^{2}\times f^{2}$ orthogonal matrix, and $\mathbf{F}$ represents $f^{2}\times f^{2}$ diagonal matrix whose elements are eigenvalues of the fourth equation in Eq.~\eqref{eq:mat}. From Eq.~\eqref{eq:mat}, since minimizing $R$ based on SVD corresponds to estimate Frobenius norm for $\mathbf{F}$ including summation of information about both known term ($\mathbf{A}$ and $\mathbf{J}$) and unknown term of $\mathbf{V}$ and zero matrix, it should be generally difficult to rewrite Eq.~\eqref{eq:mat} as Frobenius norm between a certain matrix only from known matrices information and another matrix from unknown matrices information.

With these consideration, we should introduce another strategy to modify the third term of Eq.~\eqref{eq:emrs}. 
We have previously shown that for system under pair-correlation interactions, third-order moment of a given \textit{single} pair figure is a linear function of $r$-th correlation function for PS, namely
\begin{eqnarray}
\Braket{q_{r}^{3}}\Braket{E|q_{r}}^{2} = f\left(q_{r}^{\left(\textrm{PS}\right)}\right) \Braket{E|q_{r}}^{2}.
\end{eqnarray}
This immediately suggests that using the information of PS energy ($E_{r}^{\left(\textrm{PS}\right)}$), we can change the value of $A_{rr}$ into any desired one, i.e., that can minimize the l.h.s. of Eq.~\eqref{eq:norm}. Such condition can be obtained by applying partial derivative for numerator in the last equation of Eq.~\eqref{eq:R} by $A_{rr}$, leading to the new element for $\left( r,r \right)$ element of $\mathbf{A}$:
\begin{eqnarray}
\label{eq:arr}
A'_{rr} = \pm v_{r}^{2} \pm \sum_{i\neq r} \frac{ \left( v_{i}^{2} \mp A_{ii} \right)}{3 }.
\end{eqnarray}
Therefore, we can rewrite the quadratic form and provide better approximation for SVD:
\begin{eqnarray}
&&{}^{t}\mathbf{XAX} = {}^{t}\mathbf{XA'X} + \Braket{q_{r}^{3}}\Braket{E|q_{r}}^{2} - A'_{rr}\Braket{E|q_{r}}^{2} \nonumber \\
&&\simeq \omega_{1} E_{r_{1}}^{2} + \left\{ f\left(q_{r}^{\left(\textrm{PS}\right)}\right)  - A'_{rr} \right\}\Braket{E_{q_{r}}}^{2} \nonumber \\
&&= \omega_{1} E_{r_{1}}^{2} + \frac{ f\left(q_{r}^{\left(\textrm{PS}\right)}\right)  - A'_{rr} }{ \left(q_{r}^{\left(PS\right)}\right)^{2} } \left( E_{r}^{\left(\textrm{PS}\right)}  \right)^{2},
\end{eqnarray}
where matrix $\mathbf{A}'$ has $\left(r,r\right)$ element of Eq.~\eqref{eq:arr} while all the other elemets are the same as those of $\mathbf{A}$.
When we apply the above modification for 1-NN pair correlation on fcc equiatomic system, the resultant $R=0.036029$ provide better value than $R=0.036031$ by performing only SVD where exact value of the quadratic form on $S^{f-1}$ takes 0.059608. Although in the present case, we see a slight modification for fcc 1NN pair, we can expect further modification for geometry where e.g., the first singular vector does not contain rich information about $A_{11}$. 
\begin{figure}[h]
\begin{center}
\includegraphics[width=1.00\linewidth]{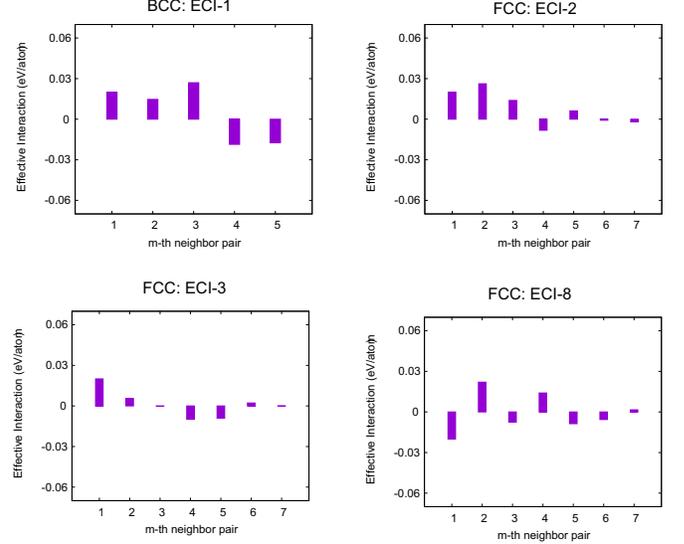}
\caption{ Prepared many-body interaction on bcc and fcc equiatomic binary system.  }
\label{fig:eci}
\end{center}
\subsection{Modification with Pade Approximation}

\end{figure}
\begin{figure}[h]
\begin{center}
\includegraphics[width=1.0\linewidth]{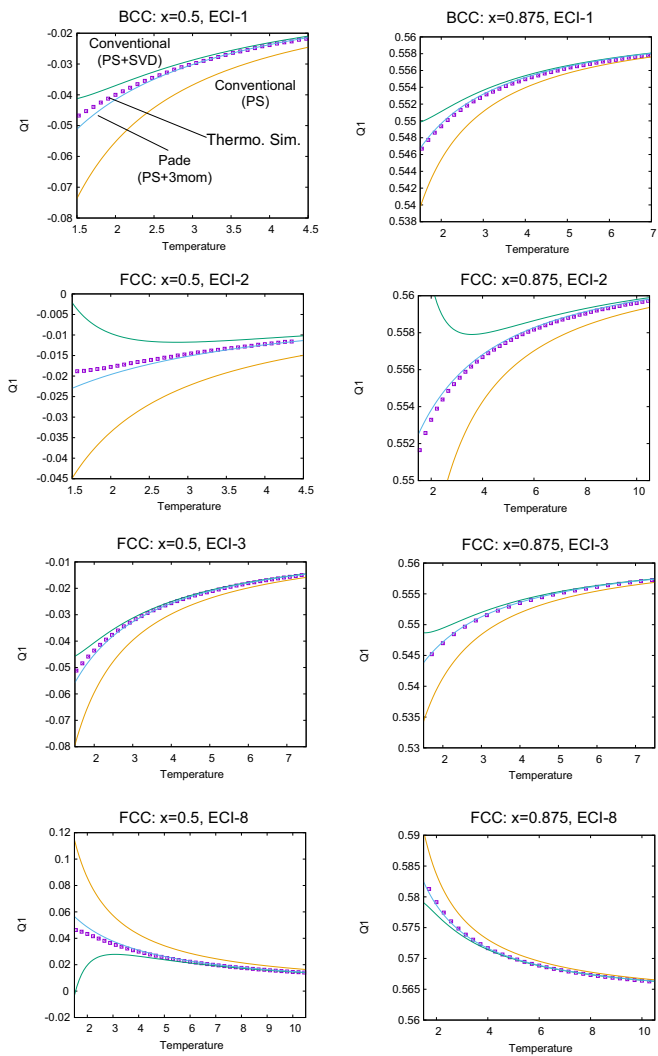}
\caption{ Temperature-dependence of the 1NN pair correlation obtained by Eq.~\eqref{eq:emrs} up to only PS energy of $E_{r}$, to further including PS2 energy of $E_{r_{1}}$, by Eq.~\eqref{eq:pade}, and by thermodynamic simulation. }
\label{fig:corr}
\end{center}
\end{figure}
Although modification based on SVD and PS energy is mathematically guaranteed for the quadratic term representation, it can strongly depend on the system, where a slight modification is found for fcc equiatomic system. Here we examine another modification for canonical average using PS and PS2 energy. Since Eq.~\eqref{eq:emrs} is a form of power series of $\beta$, we can easily expect that predictive accuracy can significantly decrease for lower temperature, which can be already seen in our previous study. 
To overcome this problem, we here perform Pade approximation of canonical average providing rational function, where $r$-th order derivative w.r.t. $\beta$ is equivalent to that for Eq.~\eqref{eq:emrs}. Corresponding representation can be immediately given by
\begin{widetext}
\begin{eqnarray}
\label{eq:pade}
\Braket{Q_{r}}_{Z} \simeq \Braket{q_{r}} - \frac{  \left( \sum_{i=1}^{g}\sqrt{\displaystyle\frac{\pi}{2}} \Braket{q_{r}}_{2} \Braket{q_{i}}_{r}^{\left(+\right)} \Braket{E|q_{i}} \right)^{2} \beta  }{ \left(\sum_{i=1}^{g} \sqrt{\displaystyle\frac{\pi}{2}} \Braket{q_{r}}_{2} \Braket{q_{i}}_{r}^{\left(+\right)} \Braket{E|q_{i}} \right) +  \left( \sum_{i,j=1}^{g}  \displaystyle{\frac{\Braket{q_{r}q_{i}q_{j}}}{2}} \Braket{E|q_{i}}\Braket{E|q_{j}}   \right)\beta   }.
\end{eqnarray}
\end{widetext}
To see the applicability of Eq.~\eqref{eq:pade}, we here artificially prepare four system on bcc and fcc equiatomic composition, where corresponding many-body interaction in terms of generalized Ising model up to 5-th (for bcc) and 7-th (for fcc) nearest-neighbor coordination is shown in Fig.~\ref{fig:eci}. 
Figure~\ref{fig:corr} shows resultant temperature-dependence of the 1NN pair correlation obtained by Eq.~\eqref{eq:emrs} up to only PS energy of $E_{r}$, to further including PS2 energy of $E_{r_{1}}$, by Eq.~\eqref{eq:pade}, and by thermodynamic simulation based on Metropois algorism. 
We can clearly see that modification based on Pade approximation can provide smaller deviation from thermodynamic average for the four prepared system, indicating that up to information about PS and PS2 energies, canonical average for structure can be well-characterized by Eq.~\eqref{eq:pade}. 
We finally note that while Eq.~\eqref{eq:pade} is expected to provide better representation for canonical average for a variety of many-body interactions, it can undergo singular when denominator of Eq.~\eqref{eq:pade} takes zero. When temperature $\beta$ satisfies such singular condition, corresponding temperature dependence should diverge: In this case, Eq.~\eqref{eq:emrs} can provide better representation for canonical average. Here the important point is that since information about PS and PS2 energies are already given, we can \textit{a priori} know whether Eq.~\eqref{eq:pade} diverges or not for considered temperature range. 

\section{Conclusions}
Based on singular value decomposition and Pade approximation, we here provide better representaion of canonical average for microscopic structure based on two specially microscopic states than that previously proposed. For the former, by choosing a element of third-order quadratic matrix to optimal value derived from information about PS energy, we show that the modified representation can always results in better approximation, while the magnitude of modification significantly depends on configurational geometry. For the latter, we demonstrate that for a variety of prepared many-body interactions, Pade approximation with PS and PS2 energy provide better representation of canonical average.

\section{Acknowledgement}
This work was supported by Grant-in-Aids for Scientific Research on Innovative Areas on High Entropy Alloys through the grant number JP18H05453 and a Grant-in-Aid for Scientific Research (16K06704) from the MEXT of Japan, Research Grant from Hitachi Metals$\cdot$Materials Science Foundation, and Advanced Low Carbon Technology Research and Development Program of the Japan Science and Technology Agency (JST).

\end{document}